\documentclass{ptephy_v1}
\preprintnumber{XXXX-XXXX}

\begin{document}

\title{Optical transmittance measurements of sheet resistors for the $\rm {CF}_4$ scintillating light in a gaseous time-projection chamber}

\author{Hiroshi Ito}
\affil{Department of Physics, Faculty of Science and Technology, Tokyo University of Science, Noda, Chiba 278-8510, Japan
\email{itoh.hiroshi@rs.tus.ac.jp}}

\begin{abstract}
A gaseous time-projection chamber (TPC) with a reconstructable $z$ coordinate for nuclear recoil tracks has been developed for dark-matter searches and $\alpha$ particle imagings in a low radioactivity background.
A TPC with a sheet-resistor field cage 
shows a potential to detect charge and photons produced by tracks
if the sheet resistor has optical transmittance for visible light,
and determine a drift length from a time difference between these signals.
In this study,
an optical transmittance of sheet resistors was measured to be
$24.5\pm0.1_{\rm stat}\pm0.6_{\rm syst}\%$
for the $\rm {CF}_4$ gas scintillating light
using an $\alpha$-particle source.
The number of photoelectrons is observed to be $\rm\sim20~p.e.$ for 5.3~MeV$\alpha$ with the existence of a sheet resistor 
in the $\rm CF_4$ gas.
Then, it is discussed how many photoelectrons to be observable by using multi-alkali-cathode phototubes and SiPMs for near-infrared light
in order to detect lower energy tracks.
\end{abstract}

\subjectindex{xxxx, xxx}
\maketitle

\section{Introduction}

A low-pressure gaseous time-projection chamber (TPC) of a low radioactivity background is a powerful devise for a direction-sensitive dark matter search and precise measurements of radioactive impurities in the material for detectors of the rare event searches.

NEWAGE \cite{NEWAGE2010, NEWAGE2015, NEWAGE2021} has been searched for weakly interacting massive particles (WIMPs) coming from the Cygnus constellation
in the events of fluorine nuclear recoil with three-dimensional tracks detected
by the gaseous TPC with $\rm CF_4$ gas filled in a ${\rm 31~cm\times 31~cm\times 41~cm}$ volume and a micro pixel chamber ($\mu$-PIC) as the readout \cite{LAuPIC2020}.
A limit on a spin dependence WIMP-proton cross section ($\sigma^{\rm SD}_{\chi-p}$) was set as 50~pb for 100~GeV/$c^2$ of WIMP mass ($M_{\rm \chi}$) (90\%~C.L.) \cite{NEWAGE2021}.
Remained background events were found to be $\alpha$ particles emitted from the $\mu$-PIC in a region between a gas electron multiplier (GEM) and the $\mu$-PIC,
i.e. the events distribute at near the $\mu$-PIC in the detector.
The current TPC has a self-trigger system, and it has an issue that the absolute $z$ position (axis along the drift) of tracks cannot be determined to reject these background.

An $\alpha$ particle imaging detector
quoted in this paper
is a low-pressure gaseous TPC,
which was refurbished from NEWAGE-0.3a detector \cite{NEWAGE2010}
in order to screen radioactive impurities on the material surface precisely for development of detectors to search WIMPs
\cite{AICHAM_2018, NIMA.aicham.2020}.
The sensitivity is evaluated to be a few $\times10^{-3}~\alpha/{\rm cm^2/hr}$
and the background event of $\alpha$ particle from radon has been found in the performance test.
Then, the detector has been used to measure
material samples 
for several experiment groups \cite{arXiv.aicham.2021}.
Most of the background events will be removed by cooled active charcoal,
however a small fraction of $\alpha$ particles from radon could remain.
The background events would be rejected if the absolute $z$ coordinate of tracks were determined in the fiducial volume.

It is known that $\rm CF_4$ gas emits scintillation light.
DMTPC \cite{DMTPC2009} has been developing an optical-readout gaseous TPC to search for the direction-sensitive WIMPs.
The demonstrator substantiated a good performance to reconstruct 3D tracks with a readout by CCD cameras as the following concept.
A nuclear recoil produces $\sim10^3$ primary electrons in 50~Torr $\rm CF_4$ gas,
the electrons drift toward an amplification field between the anode and cathode mesh,
then light is emitted by the avalanche process of the electrons
\cite{DMTPC2012,DMTPC2015}.
A limit on $\sigma^{\rm SD}_{\chi-p}$ was set to be $2.0~\rm nb$ by DMTPC for ${M_{\chi}=115~{\rm GeV}/c^2}$ (90\%~C.L.)
\cite{DMTPC2011}.

In general, conventional self-trigger TPCs are not capable of 
reconstructing an absolute $z$ coordinate of 3D tracks.
Because the trigger is issued when the electrons on the tracks arrived to the sensitive strips or pixels, a drift time cannot be measured.
A next stage of direction-sensitive WIMP search is to reconstruct the absolute $z$ coordinate for 3D tracks in the TPC.
DRIFT collaboration developed negative-ion gaseous TPC(NITPC)s based on micro patterned gaseous detectors (MPGDs) \cite{DRIFT2005}.
As the drift time is different with two negative ions (e.g., $\rm SF_6^-$ and $\rm SF_5^-$), NITPCs show a potential to determine the $z$ coordinate of tracks derived from two peaks of the signal timing.
CYGNUS collaboration has been developing a $\rm 1~m^3$ size NITPC based on $\mu$-PIC \cite{CYGNUS2020}.
The observation of $\alpha$ tracks has been established in the NITPC with a resolution of $z$ coordinate ($\sigma_z$) of $\rm 130~\mu m$
\cite{NITPC2020}.
The CYGNO project is another direction-sensitive WIMP search using $1~\rm m^3$ size optical-readout TPC.
A demonstrator has been developed consisting of $\rm 26~cm\times26~cm\times20~cm$ size TPC with an optical-readout system based on GEM and sCMOS camera \cite{CYGNO2019, CYGNO2020}.
A spatial resolution in the $z$ coordinate of
$\sigma_z=300$-$700~\rm \mu m$ was evaluated as the performance \cite{CYGNO2021}.

\begin{figure}[hptb]
    \centering
    \includegraphics[width=0.9\textwidth]
    {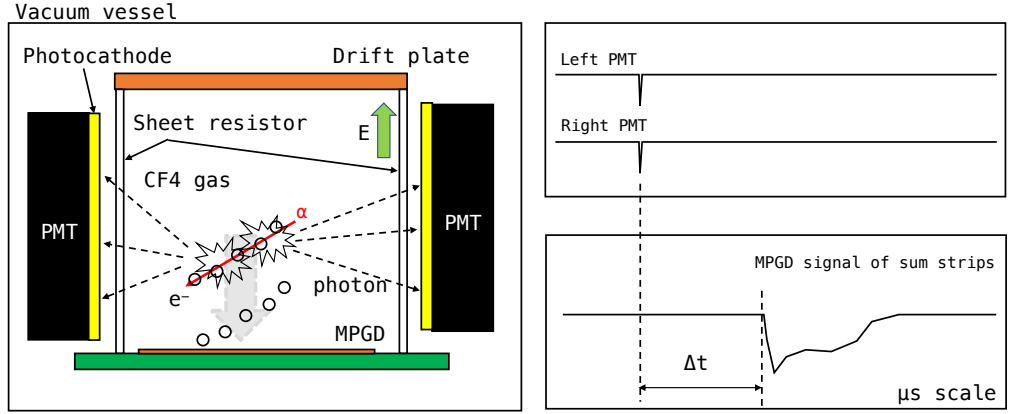}
    \caption{
    Left panel is a schematic view for both detection of electrons and photons produced by an $\alpha$ particle in a self-trigger gaseous TPC.
    The electrons (open circles) and photons (arrows with dashed line) are yield on a track of $\alpha$ particle (red arrow).
    Photons are detected by phototubes (PMT) through sheet resistor (transparent for visible light),
    then electrons move down to micro-pattern gaseous detector (MPGD) following an electric field $E$ and the charge signal is detected.
    Right panel is an example of signal waveform for PMTs and MPGD.
    The timing difference $\Delta t$ corresponds to the drift length of electrons.
    Therefore, self-trigger TPCs can reconstruct the vertex of tracks.
    }
    \label{fig:design}
\end{figure}

Recently,
an idea appeared to use sheet resistors for the field cage of the TPC, 
which makes a good linear electric field even near the wall.
A demonstrator has been developed and the performance has been evaluated \cite{miuchi.sheet-resistor.2019}.
The sheet resistor looks like a transparent for visible light,
i.e. it enables to detect the charge of tracks and photons emitted from the $\rm CF_4$ gas simultaneously.
Therefore, the $z$ coordinate of tracks would be determined from the time difference of the photon emission and the electron drift as shown in Fig.~\ref{fig:design} when both signals are detected in the TPC.
In this paper,
an optical transmittance of the sheet resistor
for scintillation light emitted from $\rm CF_4$ gas
is measured,
then a potential of the gaseous TPC using the sheet resistor for the detection of both the charge and photon signals will be discussed.

\section{Experiments}
\subsection{Setup}

Figure~\ref{fig:setup} shows a setup for this study.
A source of $\alpha$ particles was placed
in a stainless pipe-shape vacuum vessel with an inner radius of 39.4~mm.
Both ends of the vessel are connected to flanges with a quartz window.
Then, the both flanges with the window are connected to phototubes (H1161, Hamamatsu) optically.
The $\rm {CF}_4$ gas scintillating light has two wavelength regions around 200-500~nm and 500-800~nm
\cite{cf4:JINST2012, cf4:NIMA2015}.
A quantum efficiency of the phototube is $\sim25\%$ for 400~nm.
The sheet resistors are inserted between the quartz window and phototubes at both ends.
Hence, emitted light from ${\rm CF}_4$ gas in the vessel enters to a photocathode of the phototubes through the sheets.
In this study, 
an optical transmittance of the sheet resistors for the scintillation light from ${\rm CF}_4$ gas has been measured
by comparing the number of observed photoelectrons with and without the sheets.

Achilles-Vynilas is used as a sheet resistor in this study,
which has a thick of 0.2~mm,
made from polyvinyl chloride, 
a few $10^{10}~\rm \Omega/\Box$ of sheet resistivity,
and transparent for visible light.
Its radioactive impurities are very low as the upper limit
for $^{226}$Ra, {$^{228}$Ra, $^{40}$K, and $^{60}$Co
are reported to be mBq/kg level
\cite{miuchi.sheet-resistor.2019}.

\begin{figure}[!h]
    \centering
    \includegraphics[width=0.9\textwidth]
    {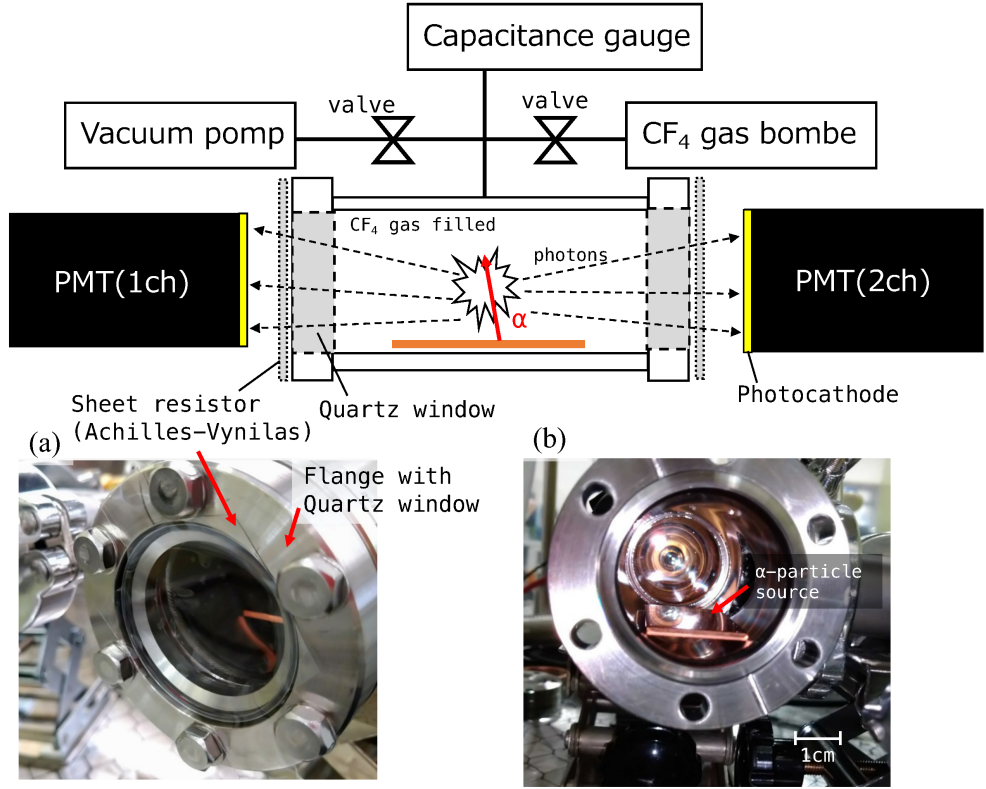}
    \caption{
    Top panel is schematic view of the experimental setup.
    Bottom left panel (a) is graphic view of the sheet resistor (Achilles-Vynilas) attached on the quartz window.
    Bottom right panel (b) is a graphic view of a copper plate as the $\alpha$ particle source placed in the vacuum vessel.
    }
    \label{fig:setup}
\end{figure}

The source is a copper plate with a size of (${\rm 2.5~cm \times 5~cm}$) and 1~mm thickness with $^{210}$Po doped to the surface.
Monoenergetic 5.3~MeV $\alpha$ particles are emitted from the surface with ${\sim \rm 0.1~s^{-1}}$ intensity.

$\rm CF_4$ gas with a purity of 99.999\% or more was used in this measurement.
The $\rm {CF}_4$ gas was filled in the vessel at 0.5 to 1.0~atm pressure,
then
the gas pressure was monitored by a capacitance gauge (M-342DG-13, Canon Anelva).
Most of the $\alpha$ particles deposit the full energy of 5.3 MeV in the gas with range of few~cm.
This setup has been kept at room temperature within $\pm1$~degree Celsius because scintillating light yield of $\rm CF_4$ tends to increase with low temperature \cite{cf4:JINST2021}.

The waveform signal of two phototubes are read by a flash ADC (DRS4, PSI) when two signals are detected simultaneously within 10~ns.
The flush ADC has 14~bits ADC to digitize signals for 1.0 Gsps sampling rate with the full range of $\rm 1024~ns$.
It is connected to a front-end board (Raspberry Pi) by USB cable, then the waveform data is stored in HDD.

\newpage
\subsection{Signal waveform}

Typical signal waveforms for a left and a right phototubes are shown in Fig.~\ref{fig:waveform}.
These correspond to a scintillation emission when the 5.3~MeV $\alpha$ particle deposited energy in the $\rm CF_4$ gas.
Both signal waveforms have a time width of 30~ns and a rise point at $\sim$320~ns in the time window.
The integrated charge within 260-400~ns is converted to 
the number of photoelectrons ($N_{p.e.}^{(j)}$),
where $j$ is a channel number of 1 and 2 for phototubes.
The total number of photoelectrons,
$N_{p.e.}^{(1)}+N_{p.e.}^{(2)}$,
for the event shown in Fig.~\ref{fig:waveform}
was derived to be 80~p.e.
Since the source is placed at the center of the vessel,
the signal amplitude detected by left and right phototubes are about the same.
Therefore, the events of light emission in the $\rm CF_4$ gas can be selected
by the existence of
both signals while the phototube dark noise events are significantly suppressed.

\begin{figure}[!h]
    \centering
    \includegraphics[width=0.8\textwidth]
    {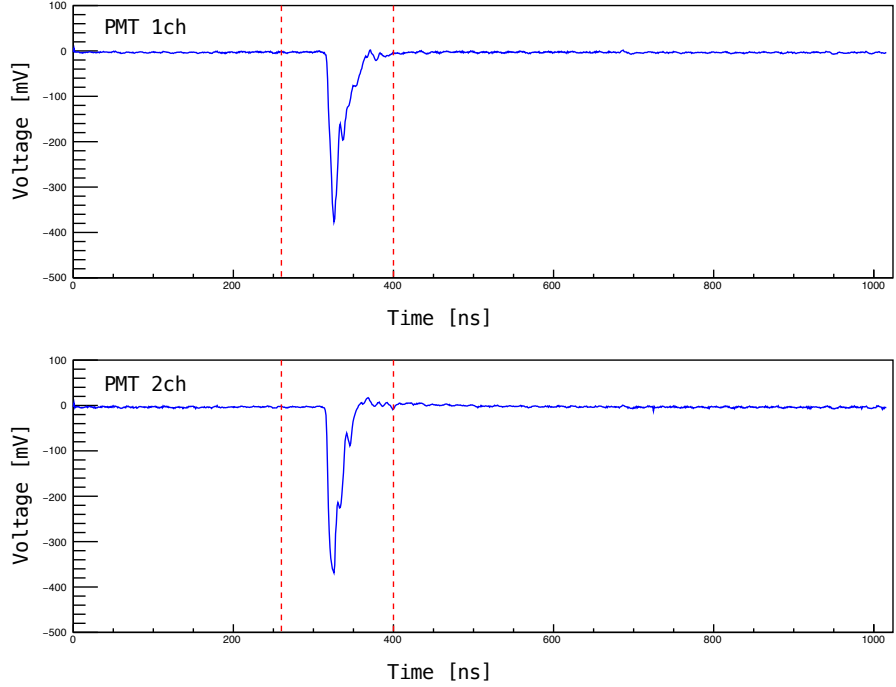}
    \caption{
    Typical signal waveform of phototubes for channel 1 (top) and channel 2 (bottom), which corresponds to scintillation emission when $\alpha$ particle deposited 5.3~MeV in $\rm CF_4$ gas.
    The integrated window is set to 260-400~ns as shown by red dashed lines.
    }
    \label{fig:waveform}
\end{figure}

\subsection{Observed photoelectrons for $\rm {CF}_4$ emission}

The relations between the number of observed photoelectrons for two phototubes with and without the $\alpha$ source are shown in Fig.~\ref{fig:relation} (a) and (b), respectively.
The gas pressure of 1.0~atm was kept in the vessel for both cases.
The live time of (a) and (b) are 15.96~hr and 112.74~hr, respectively.
An excess was found at a region of 20-80~p.e. for phototubes in comparison the data taken with and without the source.
Thus, in the setup, 
the scintillation light from $\rm CF_4$ gas 
has been detected clearly 
when $\alpha$ particles deposited energy.
In a case without the source, 
major contribution of the coincidence signals
at the region of 20-80~p.e.
are
considered
due to
Cherenkov and scintillation light in $\rm CF_4$ gas by cosmic-ray muons.

\begin{figure}[!h]
    \centering
    \includegraphics[width=0.9\textwidth]
    {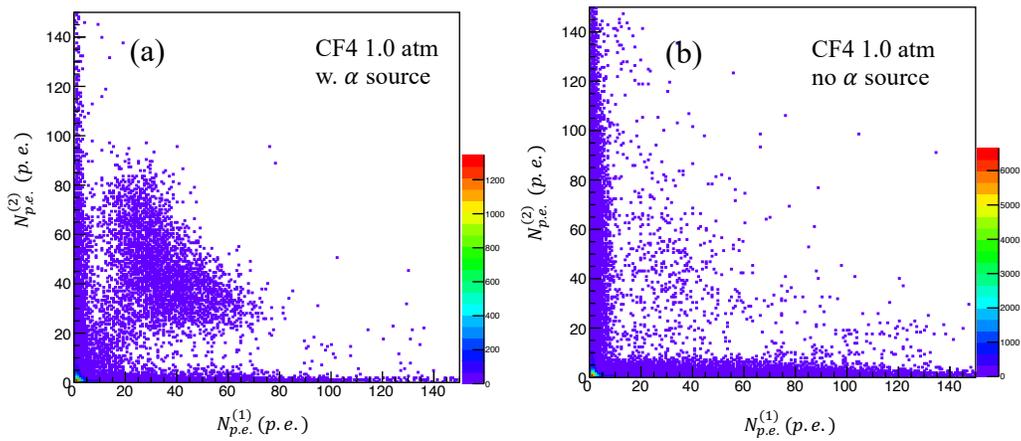}
    \caption{
    Relation between the number of observed photoelectrons from two phototubes for
    (a) with $\alpha$ particle source in 15.96~hr live time and (b) 
    without $\alpha$ particle source in 112.74~hr live time.
    }
    \label{fig:relation}
\end{figure}

\newpage
\subsection{Transmittance measurement for register sheets}

To determine the optical transmittance of the register sheets,
the measurements with/without the sheet and the source
have been performed at 1.0 atm $\rm CF_4$ gas filled in the vessel.
Figure~\ref{fig:result1} shows those histograms of count rate for the number of photoelectrons in comparison with each other,
which is normalized by the livetime in each measurement.
It is required that each phototube detects more than 2~p.e. to removed noise events.
The bottom panel of Figure~\ref{fig:result1} (a) and (b) show these residuals.
Significant differences between the clearly indicates a light emission even inserted the sheets in Fig.~\ref{fig:result1} (b).

The number of total p.e.
for the 5.3~MeV$\alpha$ deposition
is determined to be
${20.6\pm0.2}$(${81.8\pm0.2}$)
by the Gaussian fit to the residual count rate
with(without) the sheet.
As a result, 
the number of observed p.e. was reduced by the addition of the sheets, thus the transmittance is determined to be $25.2\pm0.2\%$ for the scintillation light from $\rm CF_4$ gas.

\begin{figure}[!h]
    \centering
    \includegraphics[width=0.95\textwidth]
    {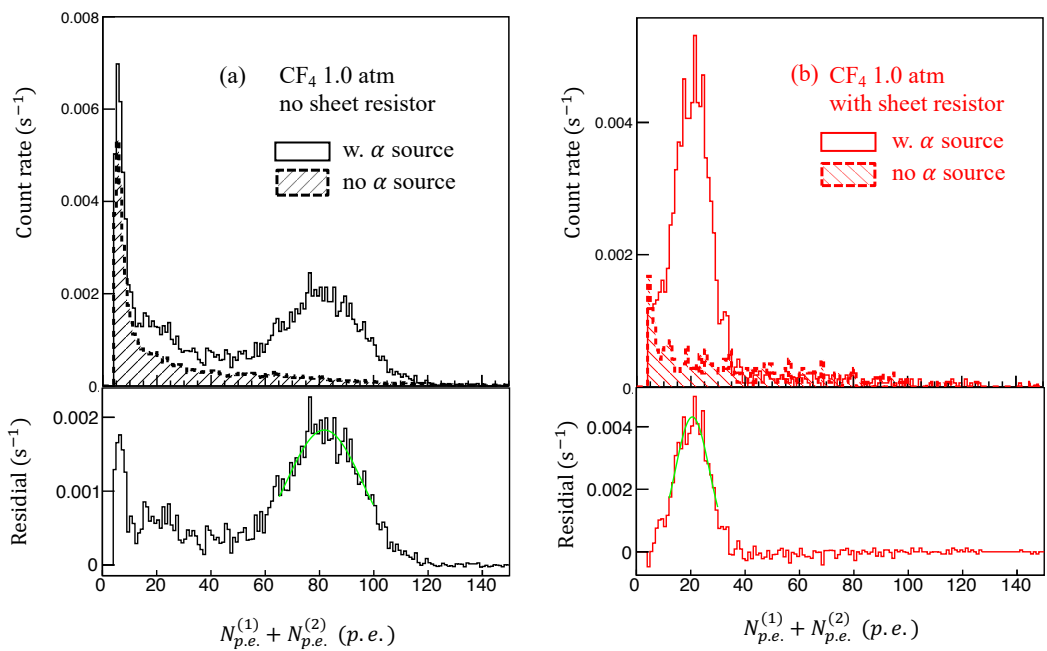}
    \caption{
    Distribution of total number of photoelectrons from the measurements for (a) with and (b) without the sheet resistor.
    The black(red) solid-white and dashed-filled histograms represent
    events with and without the $\alpha$ source and with(without) the sheet resistor at 1.0 atm $\rm CF_4$ in top panel.
    Bottom panel is the residual. 
    The green solid line is the best fit of the Gaussian function for each peak.
    }
    \label{fig:result1}
\end{figure}

As a cross check,
the measurements at 0.5~atm $\rm CF_4$ gas filled in the vessel have been also performed.
Figure~\ref{fig:result2}~(a) shows the count rate with no sheet with/without the source for 0.5~atm.
The bottom panel is the residual and the tail component at 40-70~p.e. was appeared in comparison with the spectrum of 1.0~atm.
The count rates around 60-100~p.e.
for 1.0 and 0.5~atm are calculated to be
$(5.6\pm0.5)\times10^{-2}$ and $(3.7\pm0.3)\times10^{-2}~\rm s^{-1}$
respectively.
As the gas pressure lowers from 1.0~atm to 0.5~atm, 
a path length of $\alpha$ particle gets longer from $\sim$1~cm to $\sim$3~cm.
Thus, it is considered the event rate what $\alpha$ particles cannot deposit the full energy due to a stop at the vessel wall would be increased.

The number of total p.e.
for the 5.3~MeV$\alpha$ deposition
is determined to be
${87.7\pm0.2}$ by the Gaussian fit to the residual count rate,
which is bigger than the observed p.e. for 1.0~atm.
The position of the light emission spreads along the track which gets longer in a lower pressure.
As the source plate makes be as a shadow for light toward photocathodes at both ends by itself,
the optical focus efficiency could be improved for the emission at a position far from the source.
The transmittance of the sheets would be stable for any photon amounts.
An effect what the path length is different is considered to be negligibly small to the transmittance measurement.
However, a systematic uncertainty due to the gas pressure control within $\pm2\%$ could be not negligible in comparison data with/without the sheets.

The number of total p.e.
for 5.3~MeV$\alpha$ deposition
is determined to be
${21.3\pm0.1}$
with the sheet 
as shown in Fig.~\ref{fig:result2} (b).
As a results,
the transmittance at 0.5 atm $\rm CF_4$ is determined
to be 
$24.3\pm0.1\%$,
which is smaller than that of 1.0~atm $\rm CF_4$.

A systematic uncertainty in this study is calculated
as a variance of the results obtained at two conditions,
therefore it is determined to be $\pm0.6_{\rm syst}\%$ for the average transmittance.
As a result, these values are merged to be
$24.5\pm0.1_{\rm stat}\pm0.6_{\rm syst}\%$
with an error-weighted average.

\begin{figure}[!h]
    \centering
    \includegraphics[width=0.95\textwidth]
    {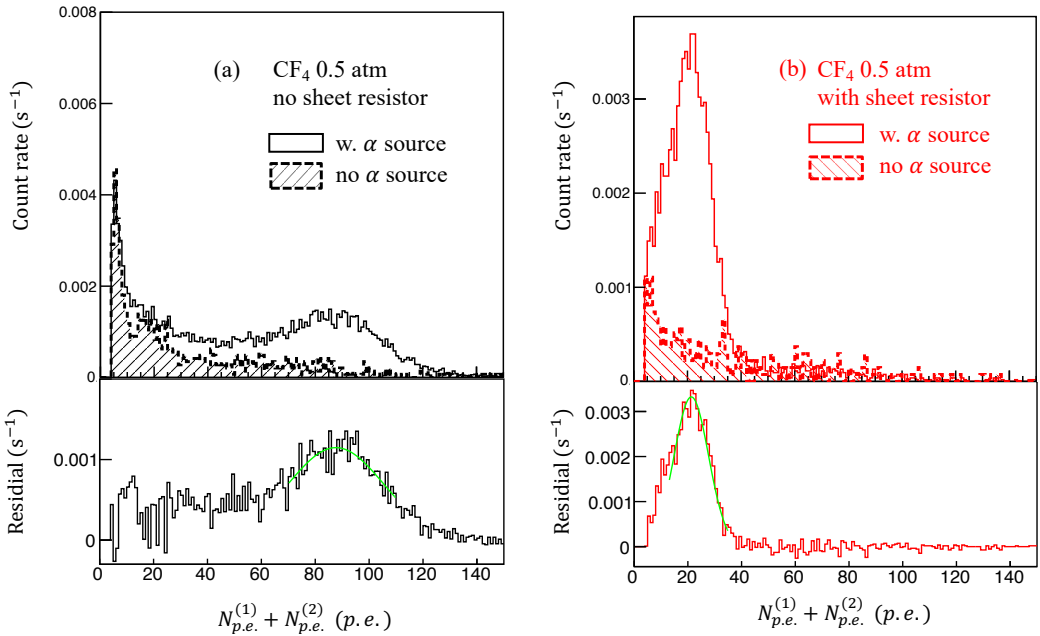}
    \caption{
    Distribution of total number of photoelectrons from the
    measurements.
    The black(red) solid-white and dashed-filled histograms represent
    events with and without the $\alpha$ particle source and with(without) the sheet resistor
    at 0.5~atm $\rm CF_4$.
    Bottom panel is the residual. 
    The green solid line is the best fit of the Gaussian function for each peak.
    }
    \label{fig:result2}
\end{figure}

\newpage
\section{Discussion}

In these measurements, the scintillation light from $\rm CF_4$ gas passing through the quartz windows and the sheet resistors is observed by standard (bi-alkali-cathode) phototubes.
Based on the measurements, 
amount of yield from $\rm CF_4$ gas for the 5.3~MeV$\alpha$ deposition and 
a potential to detect the near infrared (NI) light are discussed.

Figure~\ref{fig:discuss1} shows spectra for
a transmittance of the quartz window (blue),
a quantum efficiency of used phototubes (red dashed),
a light emission from $\rm CF_4$ gas \cite{cf4:NIMA2015} (black),
and observable photoelectrons without sheet resistors (green shade).
The left y-axis represents the transmittance of the quartz windows and sheet resistors and the quantum efficiency of phototubes which is available from datasheet in Hamamatsu \cite{Hamamatsu}.
The $\rm CF_4$ gas scintillation spectrum is normalized to 100 for the peak at 600-700~nm.
The right y-axis is a number of p.e. or photons for wavelength in comparison with the $\rm CF_4$ gas scintillation yield.
Most of the observable photoelectrons are distributed to 300-400~nm,
thus the transmittance of the sheet resistor is determined for the wavelength of blue light, as marked with a black dot.
The detection efficiency is estimated to be
$\sim5.4\%$ with the quartz window and the phototubes,
here it should be noted the optical focus efficiency is not considered.
Since $\sim$80~p.e. is obtained using phototubes without the sheet resistor,
it is estimated that at least 1480~photons are emitted from $\rm CF_4$ gas for 5.3~MeV $\alpha$ particle deposition.

\begin{figure}[!h]
    \centering
    \includegraphics[width=0.9\textwidth]
    {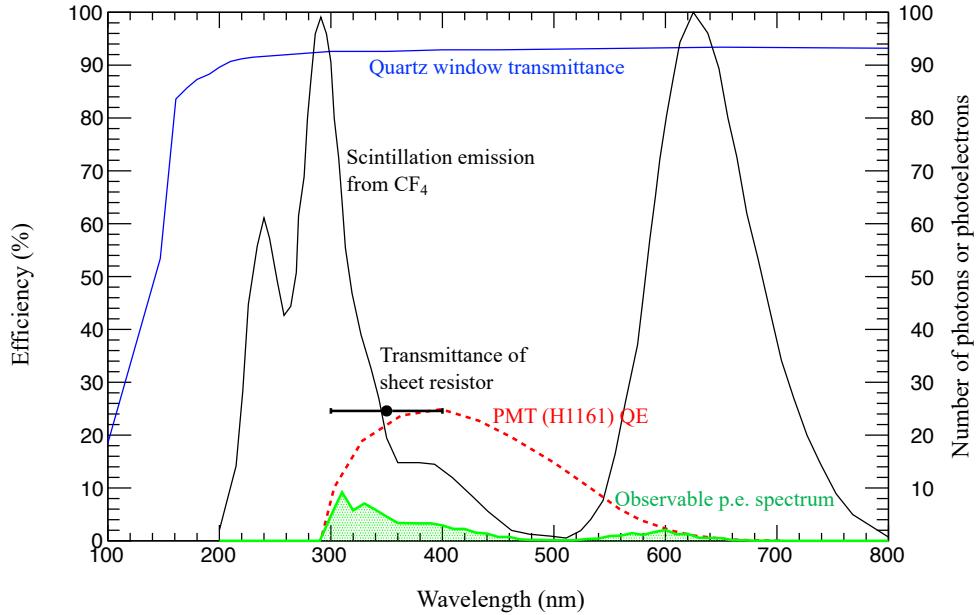}
    \caption{
    Spectra for
    transmittance of quartz window (blue),
    quantum efficiency of used phototubes (red dashed),
    light emission from $\rm CF_4$ gas \cite{cf4:NIMA2015} (black),
    and observable photoelectrons without sheet resistors (green shade).
    The black dot is measured transmittance of the sheet resistor.
    }
    \label{fig:discuss1}
\end{figure}

\newpage
Using the sheet resistor and standard phototubes (with bi-alkali cathode),
$\sim$20~p.e. for the 5.3~MeV$\alpha$ deposition is enough to observe $\alpha$ particle imagings in a low background radioactivity,
however it is too small to search a dark matter,
e.g. it is required to detect tracks of low energy nuclear recoil ($\rm \sim100~keV$).

Since the sheet looks transparent for visible light, photo sensors with a sensitivity for the NI light (500-800~nm) can be used as a possible solution.
Figure~\ref{fig:discuss2} shows a quantum efficiency of multi-alkali-cathode phototubes (red dashed), a detection efficiency of SiPMs for the NI light (magenta dashed), predictions of observable photoelectrons for multi-alkali-cathode phototubes (green shade) and SiPMs (yellow shade).
These efficiencies for the NI light sensors are available from datasheet in Hamamatsu \cite{Hamamatsu}.
The expected number of observable p.e. for multi-alkali-cathode phototubes (NI SiPMs) are calculated to be 120 (205) for the 5.3~MeV~$\alpha$ deposition in the sheet-resistor TPC, 
assuming the transmittance of the sheet resistor for 400-800~nm and at less than 400~nm
to be 90\% and 25\%,
respectively.

\begin{figure}[!h]
    \centering
    \includegraphics[width=0.9\textwidth]
    {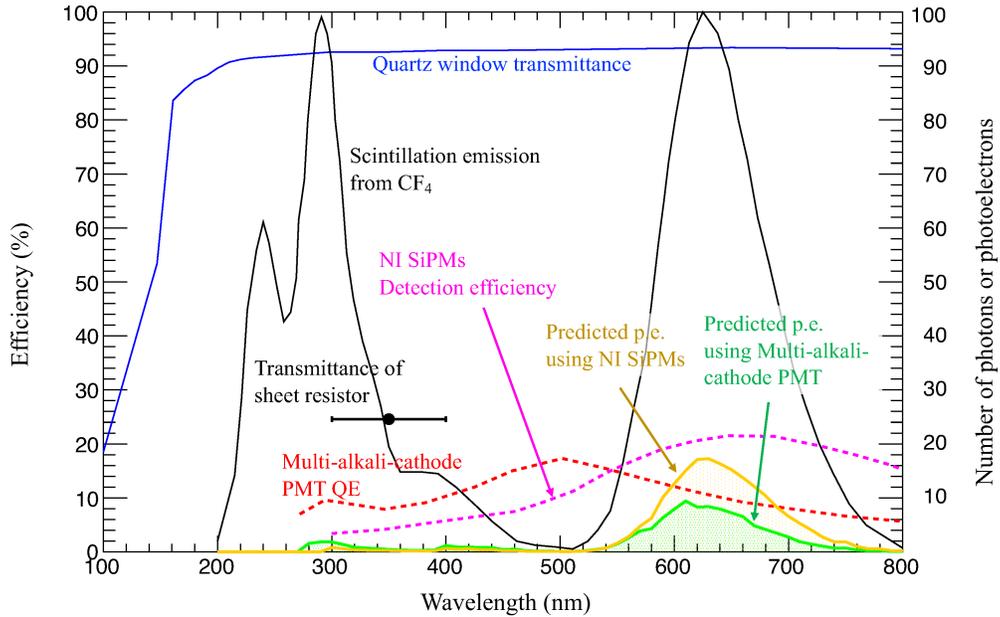}
    \caption{
    Spectra for
    transmittance of quartz window (blue),
    light emission from $\rm CF_4$ gas \cite{cf4:NIMA2015} (black),
    quantum efficiency of multi-alkali-cathode phototubes (red dashed),
    detection efficiency of SiPMs for NI light (magenta dashed),
    predicted photoelectrons for multi-alkali-cathode phototubes (green shade),
    and predicted photoelectrons for SiPMs (yellow shade).
    The black dot is result for transmittance of the sheet resistor.
    }
    \label{fig:discuss2}
\end{figure}

As a prospect,
radioactive impurities for materials of the SiPM package
must be reduced to use the SiPMs for gaseous TPCs searching a dark matter.
It was reported $\alpha$ particle emission 
from the both SiPM surfaces 
is $\rm \sim 1~h^{-1}$ intensity 
\cite{arXiv.aicham.2021}.
Furthermore, as the dark rate is not negligible for detection of low energy events, the SiPMs would be cooled down to reduce it.

\section{Conclusion}
In this work,
the optical transmittance of the sheet resistor (Achilles-Vynilas)
was measured to be 
$24.5\pm0.1_{\rm stat}\pm0.6_{\rm syst}\%$
for scintillation light from $\rm CF_4$ gas.
Further, a cross check was performed at different gas pressure of 1.0 and 0.5 atm 
and $\sim$20~p.e. was observed for the 5.3~MeV $\alpha$ particle deposition using standard phototubes and the sheet resistor.
The result indicates a good performance for $\alpha$ particle imaging detector based on the TPC with detection of both charge and photons from the same particle,
however the number of observed photoelectrons was unsatisfactory to detect low energy tracks.
In the discussion, a detection of the NI light shows a potential to enhance observable photoelectrons up to $\sim200$~p.e. for the 5.3~MeV$\alpha$ deposition.
In a next plan,
the demonstrator TPCs with the signal detection for both charge and photons would be developed using the NI light sensors.

\section*{Acknowledgment}
This work is supported by
JSPS KAKENHI Grant-in-Aid for Scientific Research, No. 20H05246.
Author would like to thank Prof. M.~Ishitsuka, Tokyo University of Science, Japan and Prof. K.~Miuchi, Kobe University, Japan for comments and discussion to improve the manuscript.

\end{document}